\documentclass[preprint,12pt]{elsarticle}
\usepackage{graphicx}  
\usepackage{amsmath} 
\usepackage{amssymb}   
\usepackage{enumerate}
\usepackage{hyperref}
\usepackage[normalem]{ulem}
\usepackage{soul,xcolor}
\setstcolor{red}
\journal{Chaos, Solitons \& Fractals}
\begin{document}
\begin{frontmatter}
\title{Aging transition under discrete time-dependent coupling: Restoring rhythmicity from aging }   
\author[mysecondaryaddress]{K. Sathiyadevi\corref{mycorrespondingauthor}}
\cortext[mycorrespondingauthor]{Corresponding author} 
\ead{sathiyadevik@gmail.com}
\author {D. Premraj$^{2}$}
\author {Tanmoy Banerjee$^{3}$}
\author {Zhigang Zheng$^{4}$}
\author {~~~~~~~~~M. Lakshmanan$^{1}$}
\address {$^{1}$ Department for Nonlinear Dynamics, School of Physics, Bharathidasan University, Tiruchirappalli 620024, Tamilnadu, India \\$^{2}$ Center for Nonlinear Systems, Chennai Institute of Technology, Chennai, India \\$^{3}$ Chaos and Complex Systems Research Laboratory, Department of Physics, University of Burdwan, Burdwan 713 104,
	West Bengal, India \\
	$^{4}$ Institute of Systems Science and College of Information Science and Engineering, Huaqiao University, Xiamen 361021, China
 }

 \begin{abstract} 
\par  We explore the aging transition in a network of globally coupled Stuart-Landau oscillators under a discrete time-dependent coupling. In this coupling, the connections among the oscillators are turned ON and OFF in a systematic manner, having either a symmetric or an asymmetric time interval.
 We discover that depending upon the time period and duty cycle of the ON-OFF intervals, the aging region shrinks drastically in the parameter space, therefore promoting restoration of oscillatory dynamics from the aging. In the case of symmetric discrete coupling (where the ON-OFF intervals are equal), the aging zone decreases significantly with the resumption of dynamism with an increasing time period of the ON-OFF intervals. On the other hand, in the case of asymmetric coupling  (where the ON-OFF intervals are not equal), we find that the ratio of the ON and OFF intervals controls the aging dynamics: the aging state is revoked more effectively if the interval of the OFF state is greater than the ON state. Finally, we study the transition in aging using a discrete pulse coupling: we note that the pulse interval plays a crucial role in determining the aging region.  For all the cases of discrete time-dependent couplings, the aging regions are shrinking and the rhythmicity gets enhanced in a controlled manner. Our findings suggest that this type of coupling can act as a noninvasive way to restore the oscillatory dynamics from an aging state in a network of coupled oscillators.  
\end{abstract}

\begin{keyword}
Aging transition \sep discrete time-dependent coupling \sep oscillation revival \sep pulse coupling



\end{keyword}

\end{frontmatter}

\section{Introduction}
\label{S:1}

\par  Complex systems abound in nature, both in the forms of man-made systems and natural systems, including power grid,  internet, social network, neural network, and so on. Depending on the network geometries, system parameters, and  interaction among the dynamical ensembles,  such complex systems exhibit  various collective behaviors, including  synchronization \cite{book1,book2,syn_zheng}, clustering \cite{clust1,clust2}, solitary state \cite{ss1,ss2,ss3,ss4}, chimera \cite{chimera1,chimera2,chimera3,chimera_zheng1,chimera_zheng2,gopal,chimera-tb}, and oscillation quenching \cite{od1,od2,od3,od4,od5,od6,od7}.  Each of these dynamical states implies a great significance with the many real-world instances.   Recent systematic analyses are therefore devoted to understanding the collective behaviors in complex systems and the topic of coupled oscillators serves as an excellent forum for such an analysis \cite{coup}. 
 
\par In the earlier stages, most researches were carried out by considering the coupling as time-independent (called static coupling).  Lately, it has been identified that, in general, most of the synaptic couplings (plasticity) in neuronal systems,    social or financial market adaptation networks, computer networks, world wide web,  and mutation processes in biological systems are time dependent \cite{app1,app2,app3}. In these systems, the interactions among the dynamical units are present only for a limited period of time. Hence the time  dependent  coupling has also been taken into account and a series of analysis carried out with random rewiring.     In the time varying network, the connectivity  among the nodes continuously varies depending on certain probability and these systems have been analyzed further spatiotemporal properties for fast and slow switching of the links of the network \cite{rw1}. The system blows up under regular ring topology for sufficiently strong coupling strength, and when some fraction of the links are randomized dynamically, this blow-up is suppressed and the system remains bounded \cite{rw2}. Recently, it has been reported that chimera states can exhibit fragility while increasing the probability of random links even for small ranges \cite{rw3}.  In the above studies, the connections among the nodes were switching continuously, but in the most realistic network, the coupling exists only until a specific task is completed and it does not exist for all other times. Thus, it is important to analyze the dynamical behaviors under transient uncoupling \cite{trans_coup1,trans_coup2}  and  ON-OFF coupling in a network of coupled oscillators. { The influence of mean-field dynamic coupling in coupled Stuart-Landau oscillators has recently been examined, which leads to the onset of various asymptotic states such as synchronization, amplitude death, and oscillation death states \cite{rev}. }
   
\par On the other hand, deterioration or degrading of dynamical activity is also a severe issue which exists in many complex networks, for instance,  cascading failure of power grids, and decreasing efficiency of living organisms are a few examples of such degradations \cite{od6,sat_ag,gow_ag}.  Originally, this was reported by Daido et al. in globally coupled oscillators \cite{ag_dai1}. The emergence of the desynchronization horn was reported while introducing a nonisochronicity parameter in globally and diffusively coupled oscillators \cite{ag_dai2}.  Further, it has been identified that the addition of asymmetry or feedback parameter enhances the dynamical robustness in an aging network \cite{ag_dg1,ag_dg2}. Effects of time-delay or random errors on the aging behavior have also been studied by varying the proportion of inactive oscillators \cite{ag_td,ag_re}.  Very recently, the aging transition in weighted homogeneous and heterogeneous networks has also been reported \cite{ag_wt}. Furthermore, the resumption of dynamism from an inactive (death) state is required for the proper functioning of many complex systems including neural networks and power grids, as inactivity may sometimes lead to improper system functioning or irrecoverable system failure \cite{rev1,rev2}. As a result, several techniques have been employed to restore the oscillations from the oscillation quenching state. In particular,  the enhancement of dynamical activity from deterioration is illustrated through low pass filtering and self-feedback factors \cite{ag_dg1,ag_dg2,lpf}.  In all of these aging related works coupling is considered to be present all the time, i.e., the coupling is static in nature and further an additional factor was added to the original system to revoke the aging state. However, the effect of aging transition on the discrete coupling is unclear and has not been explored explicitly to the best of our knowledge.   Importantly, in this paper, we show that discrete coupling can revoke aging with a  restoration of the dynamical activity from degradation. 

\par Motivated by the above, in the present work we analyze the aging transition under discrete coupling where the coupling is present (ON) for a certain duration of a time period, and for the rest of the time period it is absent (OFF). For this purpose, we consider an array of globally coupled Stuart-Landau oscillators where the coupling is periodically/aperiodically switched ON and switched OFF in discrete intervals.  We inspect the emergence of aging transition as a result of the interplay between the ratio of the active-inactive elements and the ON-OFF time of the discrete coupling scheme. We consider two types of discrete coupling, namely symmetric coupling and asymmetric coupling. In the former case, the coupling is switched ON and OFF periodically for equal intervals of time, while in the latter case the ON-OFF intervals are not equal. Significantly, we find that both types of discrete couplings suppress the aging transition and promote rhythmogenesis. In addition, we also examine the aging under pulse coupling, that is where the coupling is given as a pulse at discrete time intervals. Here also, depending upon the pulse interval, the aging region is found to be reduced.   

\par The rest of the article is organized in the following manner. The next section describes the coupling scheme along with the mathematical model of the coupled oscillator network. Section \ref{static} gives the result of the aging transition under static coupling. The results of the network under the discrete time-dependent coupling are described in Sec.~\ref{dyn}; here we systematically explore the aging transition under symmetric, asymmetric, and pulse couplings. Finally, we summarize our results and present our conclusions in Sec.~\ref{con}.     
\section{The model} 
\par To exemplify  the effect of discrete coupling, we consider a general, paradigmatic model of Stuart-Landau  limit cycle oscillator   with global diffusive   coupling,  whose governing equation can be expressed as 
\begin{eqnarray}
\dot{z_j} = f(z_j)+  \,  \chi(t)  \, \varepsilon \sum_{k=1}^{N}  ({z_k}-\, {z_j}), \quad j=1,2,...N.
\label{model1}
\end{eqnarray}
where $f(z_j) =(\lambda_j+i\omega-|z_j|^2) z_j$ and $z_j=re^{i\phi_j} = x_j+iy_j  \in C$. Here $x_j$ and $y_j$ are the state variables of the system.   $\omega$ is the eigen frequency of each oscillator and  $\varepsilon$ is the coupling strength. Further, $\lambda_j,~j=1, 2,..., N,$ are the bifurcation parameters:  $\lambda_j>0$ gives limit cycle oscillations, otherwise, steady state appears for $\lambda_j<0$.  $ \chi(t)$ is the discrete coupling function, which takes the value $0$ when the coupling is OFF and $1$ when the coupling is present (or ON). Then,  it repeats for each cycle in the entire time period.    
\begin{figure}[t!]
\centering
\hspace{-0.5cm}
\includegraphics[width=8.00cm]{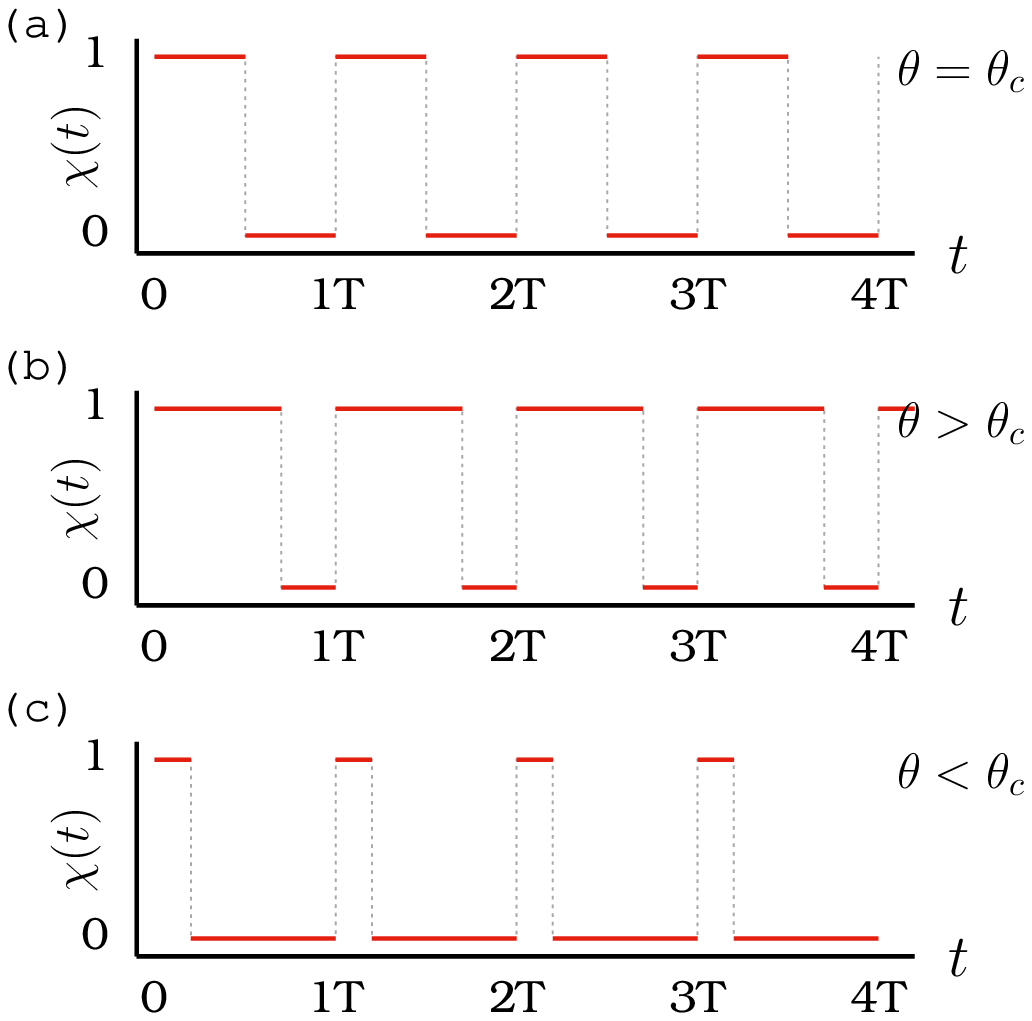}
\caption{Graphical representation of coupling function $\chi(t)$ as a function of time $(t)$. (a) Symmetric discrete coupling having an equal ON-OFF states for $\theta=\theta_c$.  The asymmetric discrete coupling consisting of (b) more ON states for  $\theta>\theta_c$ and (c) more OFF states for  $\theta<\theta_c$. }
	\label{graph}
\end{figure}
The coupling function for discrete ON-OFF coupling is defined as   
\begin{equation}
\chi(t) = 
\begin{cases}
1, &  t \in (\,lT, \,\,(l+\theta)T\,], \,\,\,\, l=0,1,2,3...  \\
0, & t \in ((l+\theta)T,\,\, (l+1)T]
\end{cases}
\end{equation}
where 
 $T$ is the time period of a cycle (i.e., ON period plus OFF period). Here, $\theta$ is the duty-cycle of the rectangular wave or ON-OFF rate which takes the values for  $0\le\theta\le1$.  $\theta=1$ denotes that the coupling function is always ON (i.e. continuous coupling), whereas $\theta=0$ denotes that the coupling is always OFF (uncoupling). Further, depending on the range of $\theta$, the discrete coupling can be classified into two types, namely symmetric discrete coupling (SDC)  and asymmetric discrete coupling (ASDC). For a more clear understanding on the classification of the discrete coupling, we depict the coupling function as a function of time in Fig.~\ref{graph}.  When $\theta=\theta_c$, the interval of ON and OFF states in every time period $T$ is equal which is referred as symmetric discrete coupling (see Fig.~\ref{graph}(a)). $\theta_c$ is the critical ON-OFF rate and  it can be taken as $\theta_c=0.5$.  If $\theta>\theta_c$ ($\theta<\theta_c$) the ON state duration is more (less) than that of the OFF state as shown in Fig.~\ref{graph}(b) and Fig.~\ref{graph}(c). Such a discrete coupling is known as an  asymmetric discrete coupling. 	
\par In order to demonstrate the aging behavior, we split the total number of oscillators into two groups: the first group is the set of active oscillators which take the values $\lambda_j = \alpha=2$, $j\in 1,2,..N(1-p)$,  while the other group is the set of inactive oscillators  having the values  $\lambda_j=-\beta = -1$,  $j\in N(1-p)+1,...N$. The parameter $p$ characterizes the ratio of inactive oscillators versus total number of oscillators.  If the ratio of inactive oscillators $p=0$, all the oscillators in the network are  active (i.e., showing oscillatory state), or else all the oscillators in the network are inactive for the inactive ratio  $p=1$ (i.e. exhibits a completely inactive or death state).   The values of inactive ratio between $0<p<1$ denotes the proportion of active and  inactive oscillators.    $N$ is the size of the network which has been  chosen as $N=500$ for our study.  To solve  Eq. (\ref{model1}), we use Runge-Kutta fourth order method with the time step of $h=0.01$. 
\section{Aging transition under continuous or static coupling} \label{static}
\par
\begin{figure}[h!]
\centering
\hspace{-0.1cm}
\includegraphics[width=10.000cm]{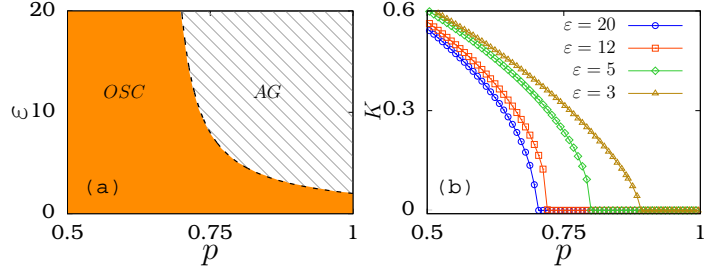}
\caption{(a) Two parameter diagram in $(\varepsilon, p)$ space for  symmetry preserving globally coupled Stuart-Landau oscillators with static coupling.  AG and OSC are the aging and oscillatory states, respectively. The dotted line  indicates the analytical critical curve for aging state.  (b) The normalized order parameter $(K)$ as a function of the ratio of inactive  oscillators  $(p)$ for different values of coupling strengths, $\varepsilon=3,~5,~12$ and $20$.    Other parameters:  $\alpha=2.0$, $\beta=-1$, $\omega=3.0$, and $N=500$.}
\label{fig1}
\end{figure}

 Before discussing the results of the discrete coupling, for a better understanding and comparison of the aging behavior, we analyze the dynamical transitions in the system (\ref{model1}) under static coupling, i.e., a constant coupling strength is   fed   for all the time. Therefore, the coupling function does not change as a function of time and it  is always in the ON state (i.e. $\chi(t)=1$).  The corresponding two-parameter phase diagram is shown in Fig.~\ref{fig1}(a) in the $(p,\varepsilon)$ space by fixing the system frequency as $\omega=3.0$.  We notice that a smaller ratio of inactive oscillators shows only oscillatory (OSC) behavior  for the entire range of coupling strength. On increasing the ratio of inactive oscillators, we observe that all the oscillators in the coupled system  (\ref{model1}) attain  the aging (AG) state at large values of coupling strength.  Further, the aging region is increased as a function of the number of inactive oscillators until all the oscillators reach the inactive state. The observed aging state is further confirmed analytically through a linear stability analysis by finding the critical ratio of inactive oscillators.  The expression for the critical ratio of inactive oscillators can be written as   $p_c= \frac{\alpha(\beta + \varepsilon)}{(\alpha +  \beta) \varepsilon}$ \cite{ag_dai1} and the details for obtaining critical ratio of inactive oscillators are given in the Appendix. The  corresponding analytical boundary for the aging state is denoted by dashed line  in Fig.~\ref{fig1}(a) and it is observed that the numerical boundary  well matches with the analytical boundary.  
\par  In order to confirm the existence of the aging state, we have estimated the normalized order parameter as a function of the inactive ratio ($p$) for four different values of coupling strength ($\varepsilon$) as shown in Fig.~\ref{fig1}(b). The order parameter $Z$ is estimated  using the relation, $Z={1}/{N} \sum_{j=1}^{N}z_j$. Then the normalized order parameter is expressed as $|Z(0)|$, $K\equiv|Z(p)|/|Z(0)|$, where  $Z(0)$ corresponds to the order parameter when the entire set of oscillators in the network contains only the active oscillators, otherwise,  $Z(p)$ is the order parameter in the presence inactive oscillators with the inactive ratio $p$. The different lines connected by points in Fig.~\ref{fig1}(b) represent the different values of the coupling strength  with $\varepsilon=3,~5,~12$ and $20$.  For all the four different values of the coupling strength, the normalized order parameter transits from finite non-zero values to a null value, indicating the transition from oscillatory to death state.   For the coupling strength  $\varepsilon=3$, it is noticed that the normalized order parameter transits from finite value to null value at a critical ratio $p=0.88$,  which indicates that the system transits from the oscillatory state to the aging state. On further increase of the coupling strength  to $\varepsilon=5,~12$ and $20$, the aging transitions occur  at  critical values of the inactive ratio $p=0.8,~0.72$ and $0.695$, respectively.  From this observation, we find an increase in the aging region with increasing coupling strength. Therefore, it is evident that the interplay of coupling strength with inactive elements is crucial in inducing the aging behavior in the considered system~(\ref{model1}).   
\par So far, we have examined the dynamical transitions in a  network of coupled oscillators with {\it continuous} or static coupling. However, as discussed earlier, in many realistic situations the coupling is task dependent, that is the coupling is discrete.   Hence it is necessary to analyze the dynamical behavior under discrete coupling.  With the above  knowledge, we demonstrate the effect of different kinds of discrete couplings on aging  behavior in the following section.    
\section{Aging transition under discrete couplings}\label{dyn}
\par In the case of discrete coupling, the coupling function is turned ON and OFF periodically for different discrete intervals of time.  In the following sections, we will be analyzing each of the discrete coupling and its consequences in detail. 
\subsection{Symmetric discrete coupling}
\par
\begin{figure}[h!]
	\centering
	\hspace{-0.3cm}
	\includegraphics[width=10.0cm]{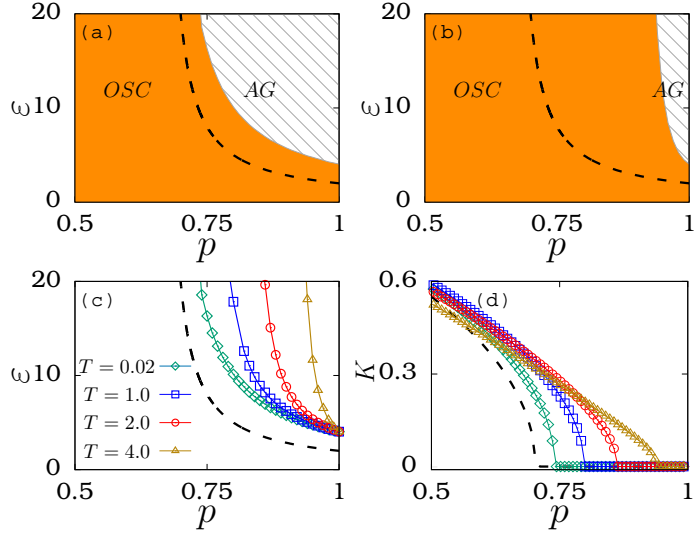}
	\caption{(a)-(b) Two-parameter diagrams in $(p,\varepsilon)$ space for time periods, where (a) $T=0.02$ and (b) $T=4.0$.  (c) The aging islands for different time periods   $T=0.02$, $T=1.0$, $T=2.0$ and $T=4.0$. (d)  Normalized order parameter as a function of critical inactive ratio, $p$, for fixed coupling strength $\varepsilon = 18.0$.  The dashed line indicates the aging boundary for continuous static coupling. Other parameters are same as in Fig.~\ref{fig1}.   }
	\label{fig2}
\end{figure}

 To understand the consequences of the symmetric discrete coupling on the  aging transition, we have plotted the two-parameter diagrams in  the $(p,\varepsilon)$ space for two different time periods, $T=0.02$ and $T=4.0$, in Figs.~\ref{fig2}(a) and \ref{fig2}(b), respectively.  The dashed line indicates the aging boundary for continuous coupling. The striped  and solid shaded regions, respectively, denote the regions for the aging and the oscillatory states. Here, we set the ON-OFF rate  $\theta=\theta_c=0.5$, i.e., half of the $T$ is in the ON state, while the other half is in the OFF state.  Figure~\ref{fig2}(a) is plotted for the small time period $T=0.02$, thereby resulting in a smaller proportion of symmetric range  in the ON-OFF coupling.   From Fig.~\ref{fig2}(a), we notice that even a smaller range of  discrete  ON-OFF coupling reduces  the spread of aging region considerably compared to the static coupling (see dashed line). We also further analyzed the aging transition by increasing the time period to  $T=4.0$. {The aging boundaries are presented in Fig.~\ref{fig2}(c) for $T=0.02$, $T=1.0$, $T=2.0$, and $T=4.0$, to demonstrate the shrinking of the aging region with increasing  discrete intervals of symmetric ON-OFF time.} We notice a gradual decrease in the aging region as the time period is increased. Increasing time period increases the  discrete intervals of symmetric ON-OFF states that causes the aging region to shrink more compared to the aging region of small discrete interval in the small time period as well as static coupling.  As a result, we find that the oscillatory region is enhanced by shrinking the aging islands, on increase in the time period $T$ of the discrete coupling. 

\par To substantiate the  observed results, the normalized order parameter is estimated as a function of the ratio of inactive oscillators for fixed coupling strength $\varepsilon=18.0$, and the corresponding plot is shown in  Fig.~\ref{fig2}(d). It is clear that the normalized order parameter transits from a finite value to  a null value i.e. the transition from  oscillatory state to aging state  occurs at the ratio of inactive oscillators	$p=0.7$ for static coupling (denoted by dashed line).  { Analogously, the normalized order parameter reaches the aging state at $p=0.74,~p=0.8,~p=0.86$, and $p=0.94$, respectively, for the time periods  $T=0.02$, $T=1.0$, $T=2.0$, and $T=4.0$. } It is clear from the observations that lengthening the ON-OFF regions leads to a widening of the oscillatory region along  with a corresponding decrease in the aging region under symmetric discrete coupling.  
\subsection{Asymmetric discrete coupling}
\par
\begin{figure} 
	\centering
	\hspace{-0.3cm}
	\includegraphics[width=10.0cm]{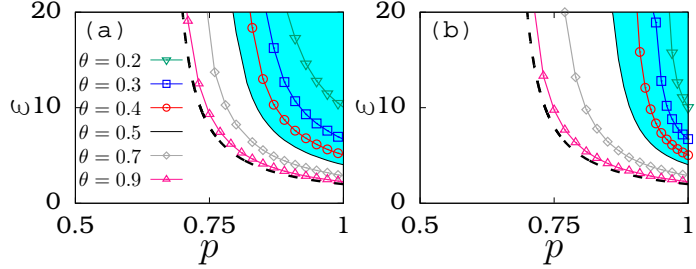}
	\caption{ The aging islands for  (a)  $T=1.0$ and (b)  $T=2.0$. The dashed line signifies the aging boundary for static coupling.  The shaded region denote the aging islands for symmetric ON-OFF states corresponding to $\theta=0.5$.  The different line connecting by points such as   downtriangles, squares, circles, diamonds, and uptriangles  designate   $\theta=0.2, ~0.3, ~0.4, ~0.7$ and   $\theta=0.9$, respectively.  Other parameters are same as in Fig.~\ref{fig1}. }
	\label{fig4}
\end{figure}

 In order to comprehend the effect of asymmetric ON-OFF coupling functions, next, we examine the role of the asymmetric ON-OFF coupling  on the aging transition. The graphical representation of asymmetric discrete coupling is shown in Figs.~\ref{graph}(b) and \ref{graph}(c).   As stated previously, the duration of the ON state is lesser than the OFF state when the ON-OFF rate is less than $0.5$, i.e. $0<\theta <\theta_c$. On the other hand, when $\theta_c<\theta <1.0$ the duration of the ON state is more than OFF state.  To explore the corresponding effects in more detail, { we have plotted the aging islands for two  different time periods  $T=1.0$ and $T=2.0$ in Figs.~\ref{fig4}(a) and \ref{fig4}(b), respectively.} The lines connected by downtriangle, square, circle, diamond, and uptriangle points in each diagram represent the ON-OFF rate $\theta=0.2, ~0.3, ~0.4, ~0.7$ and   $\theta=0.9$, respectively. The shaded region represents the case of symmetric ON-OFF rate, in which the number of the ON and OFF states are equal.  We note that the aging islands become greater than the shaded region if the ON state is more ($\theta=0.7$ and $\theta=0.9$).  If the OFF state is greater than the ON state, the aging islands are smaller than the shaded region.  Thus, the aging islands are small for $\theta=0.2, ~0.3$ and $\theta=0.4$.  It is noticeable from the observation of asymmetric discrete coupling that the aging islands are altered according to the ON-OFF rate. Thus, from the symmetric and asymmetric discrete couplings, it is revealed that the region for aging state is decreases while the restoration of the oscillatory dynamics takes place.  
\par
\begin{figure} 
	\centering
	\hspace{-0.3cm}
	\includegraphics[width=10.0cm]{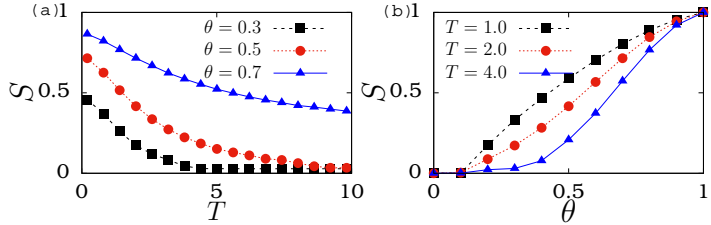}
	\caption{ Normalised area  $S=\frac{A_D}{A_S}$ as a function of (a) ON-OFF periods (T) for three different values of $\theta=0.3,~0.5$ and $0.7$, and  (b)  ON-OFF rate ($\theta$)  for three different vales of $T=1.0,~2.0$ and $4.0$. $A_S$ and $A_D$ denote  the area of the aging islands in $(p,  \varepsilon)$ parametric space for static and discrete couplings, respectively.    Other parameters are same as in Fig.~\ref{fig1}. }
	\label{area}
\end{figure}

 Furthermore, to corroborate the fact that the discrete coupling revokes the amplitude death states, we computed the normalized area ($S$) using the expression
\begin{equation}
S = \frac{A_{D} }{A_S}
\end{equation}
where $A_D$ and $A_S$ are the areas of the aging islands in the $(p,  \varepsilon)$ parametric space for discrete coupling and static coupling, respectively. Here,  the normalized area is unity for the maximum area of aging islands, and the range between $0<S<1$ represents the fact that the normalized area is lesser than that of the static coupling and that the null value denotes the complete suppression or no aging region.   Figure~\ref{area}(a) is plotted as  a function $T$ with three different values of ON-OFF rate represented by filled squares ($\theta=0.3$), circles ($\theta=0.5$) and triangles ($\theta=0.7$).
From the previous section, it is evident that the aging region is small while the ON-OFF rate is low, and that the region is larger when the ON-OFF rate is high.  As a result, the normalized aging region for $\theta=0.3$ is less than that for  $\theta=0.5$ and $\theta=0.7$ at the lower values of $T$. It is also important to note that when the ON-OFF period $(T)$ lengthens, the range of $S$ decreases, indicating that the normalized aging region shrinks. Further, we observe that for $\theta=0.3$ and $\theta=0.5$,  the normalized aging region quickly attains the null value compared to  $\theta=0.7$. Therefore, the larger aging region takes longer time to suppress the aging region compared to the shorter regions. 

\par  In addition, the normalized area is also estimated as a function of the ON-OFF rate ($\theta$) by fixing  different time periods, $T=1.0,~2.0$, and $4.0$, in Fig.~\ref{area}(b).  When the ON-OFF rate is maximum, the range $S$  takes a unit value, that is the existence of the maximum aging island.  If the   ON-OFF rate is decreased, the normalized area of the aging region gets decreased, and finally it attains the null value when the aging region completely vanishes (at $\theta=0.0$). As a result, the system shows complete oscillatory behavior.  From the observation, it is clear that the discrete coupling restores the rhythmicity in terms of ON-OFF rate or ON-OFF period.  In the following, we also investigate the restoration of rhythmicity by introducing a new type of discrete coupling called pulse coupling.  

\subsection{Pulse discrete coupling}
\par
\begin{figure}[h!]
	\centering
	\hspace{-0.2cm}
	\includegraphics[width=8.00cm]{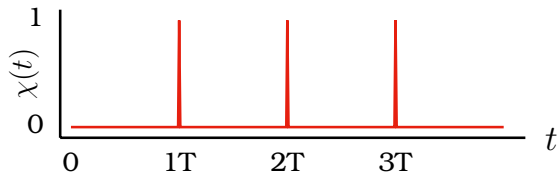}
	\caption{Graphical representation of (pulse)  coupling function $\chi(t)$ as a function of time $(t)$.     }
	\label{graph1}
\end{figure}

 Finally, we look into another interesting kind of discrete coupling, namely pulse  coupling. Comparing the above mentioned  discrete couplings,  in the pulse   coupling,  the coupling is switched ON for every $lT$ time unit once  and  in the remaining  times the coupling is in the OFF state which is clearly shown in Fig.~\ref{graph1}.  Therefore, such coupling is referred as pulse discrete coupling which is defined as 
	\begin{equation}	\chi(t) = 	\begin{cases}	1, &  if \, \,( t = lT ), \,\,\,\, l=0,1,2,3...   \\	0, & otherwise.	\end{cases}
\end{equation} 
Here 
$T$ is the pulse repetition period. 

\begin{figure}[h!]
  	\centering
  	\includegraphics[width=10.0cm]{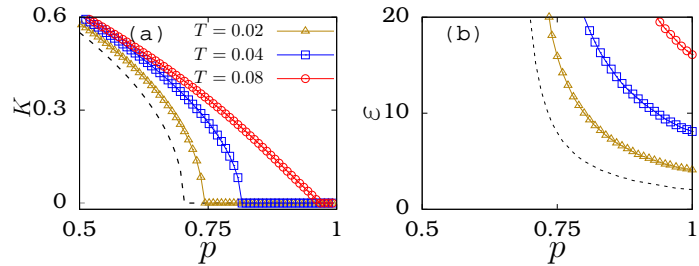}
  	\caption{(a) Normalized order paramerter for   globally coupled SL oscillators with pulse (discrete) coupling  by fixing  the  coupling strength at $\varepsilon=18.0$ and (b) the critical curves of aging state. The dashed line denotes the aging transition for continuous coupling.  Open circles, squres and triangular points represent  $T=0.02$, $T=0.04$  and  $T=0.08$,  respectively. Other parameters are same as in Fig.~\ref{fig1}. } 
  	\label{fig3}
  \end{figure}

The critical curves of the aging state for this type of coupling is shown in Fig.~\ref{fig3}.  The uptriangles,  squares and  circles connected by  lines denote the  coupling  is switched ON (pulses) instantaneously for every   $T=0.02, ~0.04$ and ~$0.08$ time intervals. The normalized order parameter for pulse discrete coupling is shown in  Fig.~\ref{fig3}(a) which illustrates that the aging transition occurs at a critical ratio of inactive oscillators $p=0.74,~0.81$, and $0.96$ for $T=0.02,~ 0.04$ and $0.08$, respectively.  From the observation, it is clear that the critical inactive ratio increases with an increase in the pulse  repetition period thereby resulting in the decreasing of the aging region.  Further,  for a more clear understanding on the effect of pulse discrete coupling, the aging boundaries are plotted in Fig.~\ref{fig3}(b) in the $(p,  \varepsilon)$ space  by fixing   $T=0.02,~ 0.04$ and $0.08$.  From the aging islands, it is clear that compared to small interval pulses, the large interval pulses decreases the aging region more effectively. Thus, it is clear that increasing the pulse repetition period reduces the aging region.   Also, a fast suppression of the aging region is observed on comparing the pulse discrete coupling with the symmetric and asymmetric  discrete coupling, that is the system quickly attains the oscillatory dynamics.  As a result, resurrection of oscillatory dynamics is observed where the coupling is given as pulses in discrete intervals.    
  
\section{Conclusions}\label{con}
\par \par Many natural and man-made systems are task-dependent in its functioning. As a result, depending on the tasks, the interactions between the dynamical ensembles are activated at specific time intervals.  Because of its importance, we explored the effect of discrete time-dependent coupling on the aging dynamics in an array of globally coupled Stuart-Landau oscillators. We have considered both symmetric (equal range of ON and OFF state) and asymmetric (OFF state is more than ON state or vice versa) coupling. 
\par We find that, under symmetric ON-OFF coupling, the aging island shrinks,
  and the oscillatory region spreads while increasing the time period of the discrete interval.  Furthermore, revoking the aging state is investigated under the asymmetric discrete coupling. In this case, also, the aging region shrinks with increasing time period. However, more significantly, here we have another control parameter to revoke the aging dynamics, namely the ratio of the ON and OFF period. The oscillatory region is enhanced if the OFF period is more than the ON period. The resumption of dynamism is validated by finding the normalized area of aging islands. Finally, we also investigated the aging transition through another kind of discrete coupling, namely pulse coupling by switching on the coupling once at every fixed time interval. We discovered that the pulse discrete coupling decreases the aging region more effectively than the symmetric or asymmetric discrete coupling.
\par From this work,  we find that discrete time-dependent couplings, including symmetric, asymmetric, and pulse couplings, favor oscillatory dynamics with the suppression of aging behavior. Therefore, our study establishes a noninvasive way of restoring rhythmicity from an aging state in a network of coupled oscillators. The discrete time-dependent coupling can be easily realized in experiments, e.g., in electronic experiments, an analog switch will serve the purpose of implementing the ON-OFF coupling \cite{tbook,sed}. Therefore, we believe that our coupling scheme can be implemented and exploited in man-made systems and natural systems also to restore oscillations from an aging state.
\section*{Appendix}\label{ap}
\par The stability of aging state for a static coupling is obtained by finding the critical inactive ratio  ($p_c$) using the linear stability analysis as discussed in \cite{ag_dai1}.  In order to find the critical inactive ratio for aging state, the oscillators in the network are split  into two groups by setting $z_j=A$ $(j\in 1,2,..N(1-p))$ for active oscillators and $z_j=I$ ($j\in N(1-p)+1,...N$) for inactive oscillators. Then the reduced model of Eq.~(\ref{model1}) can be written as
\begin{eqnarray}
\dot{A} &=& (\alpha+i \omega-|A|^2)A  + \varepsilon (1-p) (I-A), \nonumber \\
\dot{I} &=& (-\beta+i \omega-|I|^2)I  + \varepsilon p (A-I)
\label{red}
\end{eqnarray}
The Jacobian matrix of Eq.~(\ref{red}), for determining the stability of aging state is given by\\ 
\begin{center}
	$J=\begin{bmatrix}
	\alpha-\varepsilon p + i \omega & \varepsilon p\\
	\varepsilon(1-p) &  -\beta-\varepsilon(1-p) + i \omega
	\end{bmatrix}$
\end{center}
The corresponding eigenvalues from the Jacobian matrix are expressed as
\begin{eqnarray}
\lambda_{1,2}= \alpha-\beta-\varepsilon \pm \sqrt{(a + b) (a + b + 2 k - 4 k p)} + 2 i \omega
\end{eqnarray}
Further, the critical ratio of inactive oscillators $p_c$ is obtained by solving the real parts of the  above eigenvalues, which can be expressed as  
\begin{eqnarray}
p_c= \frac{\alpha(\beta + \varepsilon)}{(\alpha +  \beta) \varepsilon}
\label{pc}
\end{eqnarray} 
Using Eq.~\ref{pc}, the stability region for aging state corresponding to the static coupling is delineated in the main text. 
\section*{Acknowledgments} 
\par KS thanks the DST-SERB, Government of India, for providing National Post Doctoral fellowship under the Grant No. PDF/2019/001589. TB acknowledges the financial support from DST-SERB [CRG/2019/002632]. ZZ is partially supported by the National Natural Science Foundation of China under the Grant No. 11875135, the Quanzhou city Science \& Technology Program of China (No. 2018C085R),
and the Scientific Research Funds of Huaqiao University (Grant No. 15BS401).  ML is supported by the DST-SERB National Science Chair program. 

\section*{Compliance with Ethics Requirements}

This article does not contain any study with human or animal subjects.
%
\section*{Declaration of interests}
The authors declare that they have no known competing financial interests or personal relationships that could have appeared to influence the work reported in this paper.   

\end{document}